\documentclass{iau} 
\usepackage{graphicx}

\title[SDSS galaxies with FIRST cores] 
{Catalogue with visual morphological classification of 32,616 radio galaxies with optical hosts}

\author[Natalia \.{Z}ywucka, Dorota Kozie{\l}-Wierzbowska, \& Arti Goyal]   
 {Natalia \.{Z}ywucka$^1$,
 Dorota Kozie{l}-Wierzbowska$^2$,
 \and 
  Arti Goyal$^2$}

\affiliation{$^1$Centre for Space Research, North-West University, \\ 
2520, Potchefstroom, South Africa \\email: {\tt n.zywucka@oa.uj.edu.pl} \\[\affilskip]
$^2$Astronomical Observatory, Jagiellonian University, \\ 
30-244, Krak\'{o}w, Poland \\ email: {\tt dorota.koziel@uj.edu.pl, arti@oa.uj.edu.pl} }

\pubyear{2019}
\volume{356}  
\jname{Nuclear Activity in Galaxies Across Cosmic Time}
\editors{M. Povi\'c, P. Marziani, J. Masegosa, H. Netzer,\\ S. H. Negu, \& S. B. Tessema, eds.}
\begin{document}

\maketitle

\begin{abstract}

We present the catalogue of Radio sources associated with Optical Galaxies and having Unresolved or Extended morphologies I (ROGUE~I). It was generated by cross-matching galaxies from the Sloan Digital Sky Survey Data Release\,7 (SDSS DR\,7) as well as radio sources from the First Images of Radio Sky at Twenty Centimetre (FIRST) and the National Radio Astronomical Observatory VLA Sky Survey (NVSS) catalogues. We created the largest handmade catalogue of visually classified radio objects and associated with them optical host galaxies, containing 32,616 galaxies with a FIRST core within 3 arcsec of the optical position. All listed objects possess the good quality SDSS DR\,7 spectra with the signal-to-noise ratio $>$\,10 and spectroscopic redshifts up to $z=0.6$. The radio morphology classification was performed by a visual examination of the FIRST and the NVSS contour maps overlaid on a DSS image, while an optical morphology classification was based on the 120 arcsec snapshot images from SDSS DR\,7. 

The majority of radio galaxies in ROGUE~I, i.e. $\sim$\,93\%, are unresolved (compact or elongated), while the rest of them exhibit extended morphologies, such as Fanaroff-Riley (FR) type I, II, and hybrid, wide-angle tail, narrow-angle tail, head-tail sources, and sources with intermittent or reoriented jet activity, i.e. double--double, X--shaped, and Z--shaped. Most of FR~IIs have low radio luminosities, comparable to the luminosities of FR~Is. Moreover, due to visual check of all radio maps and optical images, we were able to discover or reclassify a number of radio objects as giant, double--double, X--shaped, and Z--shaped radio galaxies. The presented sample can serve as a database for training automatic methods of identification and classification of optical and radio galaxies.

\end{abstract}

\firstsection 
\section{Introduction}

Up to date, the majority of catalogues containing optical galaxies with radio counterparts do not include detailed morphological classification of radio structure and/or associated optical host galaxy (e.g.,  \cite[Lin et al. 2010]{Lin10}, \cite[Best \& Heckman 2012]{Best12}). We provide a catalogue of radio sources identified with optical galaxies, possessing spectroscopic redshift and good quality optical spectra from the Sloan Digital Sky Survey Data Release\,7 (SDSS DR\,7; \cite[Abazajian \etal\ 2009]{Abaz09}), measured radio flux densities of radio structures with a flux density limit of the First Images of Radio Sky at Twenty Centimetre (FIRST; \cite[White \etal\ 1997]{Whit97}), visually assigned morphological classifications of the radio structure and the optical host galaxy. The catalogue of Radio sources associated with Optical Galaxies and having Unresolved or Extended morphologies I (ROGUE~I) is the largest sample of spectroscopically selected radio galaxies, covering $\sim$\,30\% of the entire sky. 

As a parent sample, containing 662,531 unique SDSS galaxies, we used objects from the Red Galaxy Sample (\cite[Eisenstein \etal\ 2001]{Eise01}) and the SDSS Main Galaxy Sample (\cite[Strauss \etal\ 2002]{Stra02}), introducing a limit on a signal-to-noise ratio in the continuum at 4020 \r{A} $\geq$ 10 (\cite[Kozie{\l}-Wierzbowska \& Stasi{\'n}ska 2017]{Kozi17}). In order to identify the SDSS galaxies with radio sources, we cross-match the positions of the SDSS galaxies with the sources from the FIRST catalogue by applying a matching radius of 3 arcsec. Subsequently, we generated the FIRST and the NRAO VLA Sky Survey (NVSS; \cite{Cond98}) radio contour --- optical gray images, centred at the host galaxy position and having an angular size equal to a linear size of 1 Mpc at the source distance. This allowed us to visualise the morphologies of small and giant radio sources as well as associate them with an SDSS galaxy. Both radio sky surveys used in this work were conducted at 1.4 GHz and have different angular resolution of the radio images and of the sensitivity for the point-like and extended/diffuse emission. FIRST provides 5.4 arcsec synthesized beam size images and is complete down to 1 mJy flux density limit for point-like sources, while for NVSS these parameters are 45 arcsec and 2.5 mJy, respectively. Using this procedure, we have found 32,616 matching sources, which we further classified morphologically. 629,815 remaining galaxies from the parent sample can still host extended radio emission without a core. This will be searched for in future work on \textit{ROGUE II: A catalog of SDSS galaxies without FIRST cores}.


\section{Classification schemes}

Our radio morphological classification scheme is detailed and complex, consisting of:
\begin{itemize}
\item compact, i.e. point-like single-component sources,
\item elongated, i.e. elliptical profile single-component sources, 
\item FR~I, FR~II, and hybrid, i.e. linear structure brighter near core, linear structure brighter near edges, and one lobe of FR~I and another of FR~II morphology, respectively, 
\item Z--shaped, i.e. sources with Z-- or S--shaped radio morphology, 
\item X--shaped, i.e. sources with X--shaped radio morphology,
\item double-double, i.e. two pairs of collinear lobes, 
\item narrow-angle, wide-angle, and head-tail, i.e. sources with angle between lobes $<90^{\circ}$, with angle between lobes $>90^{\circ}$, and with bright core and a tail, respectively,  
\item O I and O II, i.e. one-sided sources with FR~I or FR~II lobe, 
\item halo, i.e. diffuse radio emission around the core,
\item star-forming region, i.e. emission from the host galaxy
\end{itemize}
We also classified some of the sources as not clear (radio source with unclear morphology), blended (radio emission blended with other source), and not detected (when an optical galaxy is not a host of the radio emission).\\

We used the standard Hubble classification scheme to assign a morphological type of the optical host galaxy, extending it with some additional types:
\begin{itemize}
\item spiral, i.e. disc galaxy with visible spiral arms,
\item elliptical,
\item lenticular, i.e. disc galaxy without spiral arms,
\item distorted,
\item ring, i.e. ring-like shape,
\item merger, mainly major merger,
\item star-forming region, i.e. SDSS spectrum of star-forming region, not galaxy center,
\item off-center, i.e. spectrum not corresponding to star-forming region.
\end{itemize}
We also included in our scheme an interacting galaxy (with visible signs of interaction) and barred galaxy (spiral or lenticular with prominent bars). Sources with uncertain attribution of any of aforementioned radio or optical types are marked with \textit{p} for \textit{possible}.

\section{Results}

Our cross-matching procedure and visual classification allowed us to find that:
\begin{itemize}
\item unresolved (compact and elongated) radio sources dominate in ROGUE~I, constituting 92\%. The remaining 8\% of sources show extended morphology,
\item secure and possible radio sources of FR~I, II, and hybrid form group containing 73\% of the extended sources, bent (wide-angle, narrow-angle, head-tail) sources are 23\%, and sources with intermittent or re-oriented jet activity (double--double, X--shape, Z--shape sources) --- 3\%,
\item we discovered or reclassified 55 giant, 16 double--double, 9 X--shaped, and 25 Z--shaped radio galaxies, 
\item the optical morphological classification revealed that 64\% of radio galaxies in ROGUE~I have elliptical, 15\% spiral, 12\% distorted, and 7\% lenticular hosts. 
\end{itemize}

\section{Summary and conclusions}

We have created the largest handmade catalogue of 32,616 radio sources associated with optical galaxies. All ROGUE~I objects have provided 1.4 GHz radio flux densities of the core from FIRST and flux densities of the total emission from FIRST or NVSS, radio and optical morphological classifications, luminosity distance, spectroscopic redshifts, good quality spectra, and apparent optical magnitudes from SDSS. The ROGUE~I sample can serve as a database for training automatic methods of radio sources' identification and their morphological classification.

\acknowledgements{N\.Z work is supported by the NCN through the grant DEC-2014/15/N/ST9/05171. DKW acknowledges the support of Polish National Science Centre (NCN) grant via 2016/21/B/ST9/01620, and AG acknowledges support from the Polish National Science Centre through the grant 2018/29/B/ST9/02298.}


\begin{thebibliography}{}

\bibitem[Best \& Heckman 2012]{Best12}
{Best, P. N., \& Heckman, T. M.,} 2012, 
\textit{MNRAS}, 421, 1569

\bibitem[Lin \etal\ 2010]{Lin10}
{Lin, Y.-T., Shen, Y., Strauss, M. A., Richards,G. T., \& Lunnan, R.} 2010, 
\textit{ApJ}, 723, 1119

\bibitem[Abazajian \etal\ 2009]{Abaz09}
{Abazajian, K. N., Adelman-McCarthy, J. K., Ag{\"u}eros, M. A., et al.,} 2009, 
\textit{ApJS}, 182, 543

\bibitem[White \etal\ 1997]{Whit97}
{White, R. L., Becker, R. H., Helfand, D. J., \& Gregg, M. D.,} 1997, 
\textit{ApJ}, 475, 479

\bibitem[Condon \etal\ 1998]{Cond98}
{Condon, J. J., Cotton, W. D., Greisen, E. W., et al.,} 1998, 
\textit{AJ}, 115, 1693

\bibitem[Eisenstein \etal\ 2001]{Eise01}
{Eisenstein, D.~J., Annis, J., Gunn, J.~E., et al.,} 2001,
\textit{AJ}, 122, 2267

\bibitem[Strauss \etal\ 2002]{Stra02}
{Strauss, M.~A., Weinberg, D.~H. \& Lupton, R.~H.,} 2002, 
\textit{AJ}, 124, 1810

\bibitem[Kozie{\l}-Wierzbowska \& Stasi{\'n}ska 2017]{Kozi17}
{Kozie{\l}-Wierzbowska, D. \& Stasi{\'n}ska, G.,} 2017, 
\textit{MNRAS}, 415, 1013

\end{thebibliography}
\end{document}